\documentclass[runningheads]{svmult}

\usepackage{graphicx}  
\usepackage{physprbb}  

\newcommand{\greeksym}[1]{{\usefont{U}{psy}{m}{n}#1}}
\newcommand{\umu}{\mbox{\greeksym{m}}}

\begin{document}

\title*{MAXIMUS: Exploiting the Full Power of OzPoz}

\toctitle{MAXIMUS: Exploiting the Full Power of OzPoz}

\titlerunning{MAXIMUS}

\author{Matthew Colless\inst{1} \and Keith Taylor\inst{2}}

\authorrunning{Matthew Colless \and Keith Taylor}

\institute{Research School of Astronomy \& Astrophysics, The Australian
           National University, Weston Creek, ACT 2611, Australia
\and       Dept of Astronomy, California Institute of Technology,
           MS105-24, Pasadena, CA 91125, USA}

\maketitle

\begin{abstract}
We propose a new multi-object spectrograph for the VLT. MAXIMUS (MAXimum
MUltiplex Spectrograph) will fully exploit the multiplexing capabilities
of the OzPoz fibre positioner in order to extend and complement FLAMES
and VIMOS in covering observational parameter space, and to meet the
increasing demand for multi-object spectroscopy by ESO users in the next
decade.
\end{abstract}

\section{The OzPoz fibre positioner}

The OzPoz fibre positioner is being built by the AUSTRALIS consortium
(consisting of the Anglo-Australian Observatory, the Australian National
University and the University of New South Wales) for the FLAMES
fibre-spectroscopy facility on the VLT ({\tt
http://www.eso.org/instruments/flames/OzPoz.html}).

OzPoz allows full access to the 25\,arcmin diameter (0.136\,deg$^2$)
field of view of the VLT Nasmyth focus. It offers four field-plate
positions, only two of which are used by GIRAFFE and UVES -- the other
two positions are currently unused. Up to 600 single-object fibres can
be mounted on each field-plate, and the positioning robot is capable of
configuring all 600 fibres in 50 minutes. As well, OzPoz handles
multiple deployable fibre integral field units (d-IFUs), which can
be as numerous as detector area allows.

\section{Exploiting OzPoz}

There are two main ways in which the capabilities of OzPoz could be
fully exploited by new VLT instrumentation.

One way is essentially an \textbf{upgrade} option for FLAMES, leaving
the existing feeds to GIRAFFE and UVES spectrographs in place and
utilising the two free field plates to feed a new spectrograph. Each of
the two new field plates would hold 560 single 1~arcsec diameter fibres
and also 15--20 d-IFUs each with 127 0.3~arcsec fibres giving a
4~arcsec diameter field.

The other option is a full-fledged \textbf{second-generation instrument}
for the VLT, and would use all four of OzPoz's field plates. In this
case one would have more flexibility in the mix of single fibres and
d-IFUs. For example, one might chose to have two plates with 600 single
1~arcsec diameter fibres and two plates with 60--80 d-IFUs with 2~arcsec
diameter fields ($37\times0.3$~arcsec); many other arrangements
(particularly for the d-IFUs) are also possible.

\begin{figure}
\begin{center}
\includegraphics[width=0.67\textwidth]{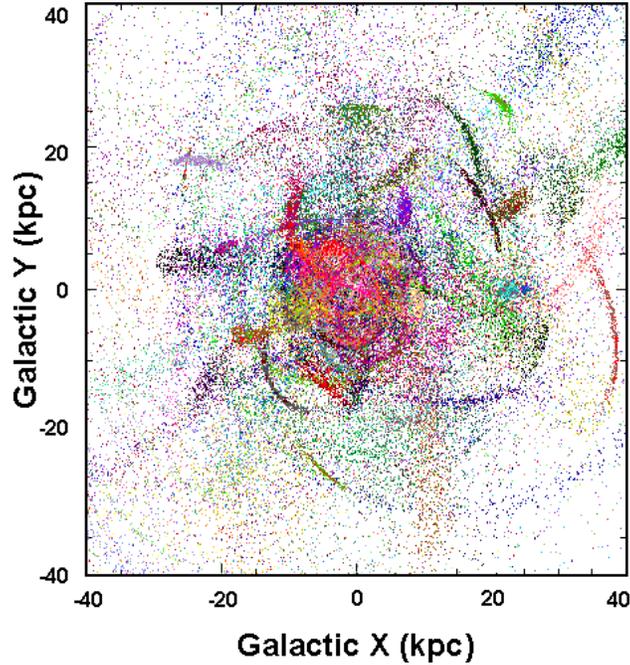}
\end{center}
\caption[]{Simulation of the tidal streams of stars accreted into the
Galactic halo.}
\label{fig:spaghetti}
\end{figure}

\section{Science Drivers for MAXIMUS}

\begin{figure}
\begin{center}
\includegraphics[width=\textwidth,angle=0]{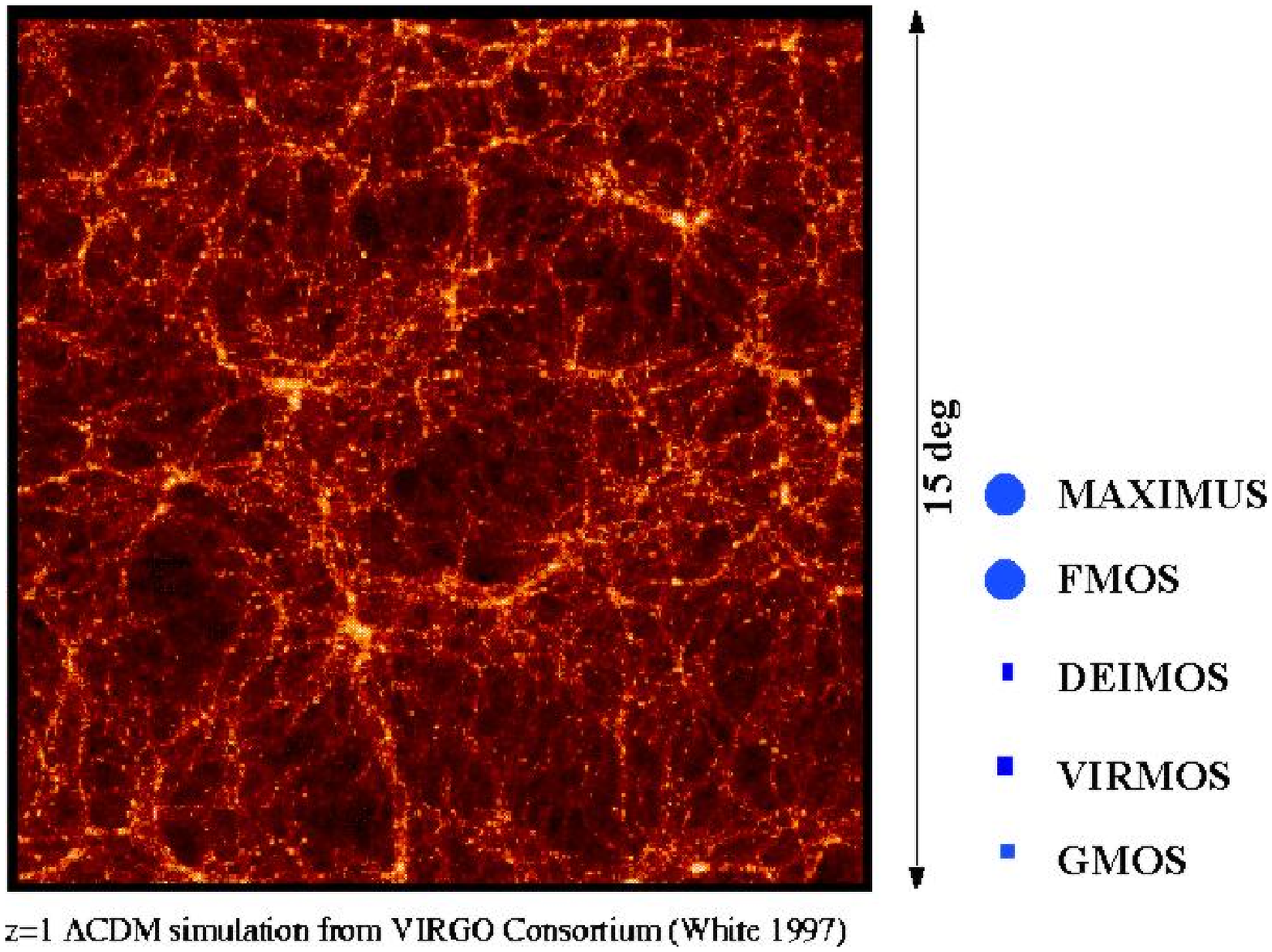}
\end{center}
\caption[]{Simulation of the large-scale structure at a redshift $z=1$
by the Virgo Consortium [2], and the relative fields of view of various
multi-object spectrographs.}
\label{fig:highzlss}
\end{figure}

The scientific demand for multi-object spectrographs is so strong that
every single 8-metre telescope (not just every 8-metre observatory) will
have at least one: Gemini North and South both have a GMOS (also
FLAMINGOS-II); Keck I+II will have LRIS multislits, DEIMOS and KIRMOS;
MMT will have Hectospec; Subaru will have FMOS. The four VLT unit
telescopes will have FLAMES, VIMOS and NIRMOS. The new deep, wide-field
imaging surveys in the visible, UV, IR, X-ray and radio will only
increase the demand for deep, wide-field, multi-object spectroscopy on
8-metre telescopes. Spectroscopic follow-up of surveys carried out with
VISTA will be particularly important to VLT users.

With such a plethora of multi-object spectrographs, it is important for
new instruments to have distinguishing advantages. MAXIMUS aims to
provide VLT users with the maximum multiplex advantage allowed by the
OzPoz fibre positioner, for both unresolved single-fibres and small,
deployable integral fields. The science drivers are observations that
require either: (i)~the highest possible multiplex gain for unresolved
single-object spectroscopy over the widest possible field of view, or
(ii)~high-multiplex resolved spectroscopy over a few square arcseconds
with seeing-limited sampling over the widest possible field of view.

An example of a Galactic programme driven by a wide field and maximum
multiplex is the attempt to recover the merger history of the Galaxy's
halo by mapping the tidal streams of the accreted stars.
Figure~\ref{fig:spaghetti} shows a simulation of these stellar streams
in what has been aptly referred to as the spaghetti model for halo
formation [1]. It is immediately apparent that only a densely-sampled
wide-field survey of stellar velocities (requiring a high-multiplex,
wide-field spectrograph with good spectral resolution) will be able to
unravel the complex structure of the halo.

Similar instrumental requirements are demanded by surveys of the
evolution of large-scale structure at high redshift.
Figure~\ref{fig:highzlss} shows the fields of view of multi-object
spectrographs on 8-metre telescopes compared to a simulation of the
large-scale structure in the galaxy distribution at a redshift of $z=1$
[2]. Again, it is readily apparent that a high-multiplex, wide-field
spectrograph will be essential for mapping these structures on the
scales of interest.

\section{Conceptual designs for MAXIMUS}

We propose two possible conceptual designs for MAXIMUS:

\begin{figure}
\vspace*{1cm}
\begin{center}
\includegraphics[width=\textwidth]{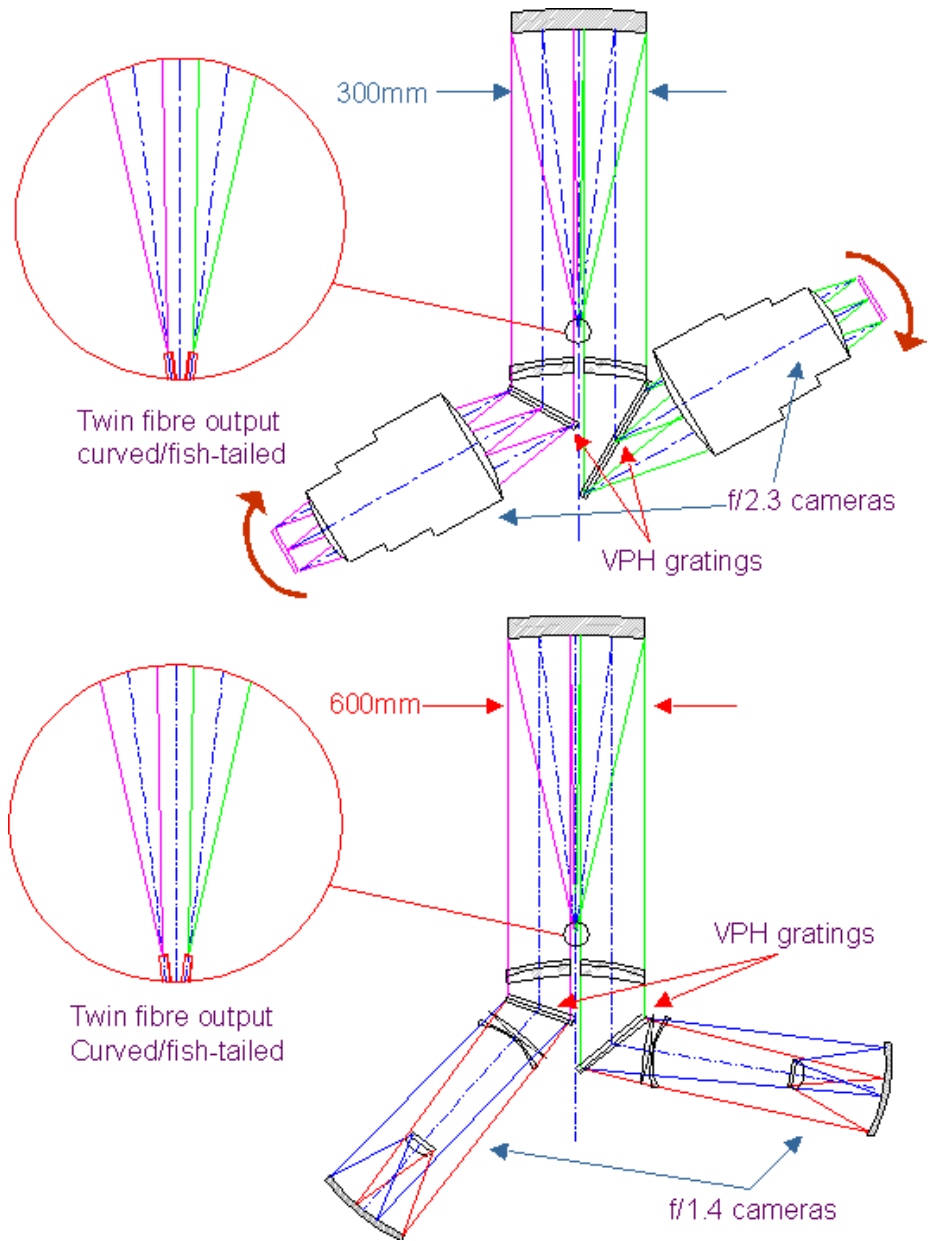}
\end{center}
\caption[]{Conceptual designs for MAXIMUS -- design~\#1 above and
design~\#2 below.}
\label{fig:designs}
\end{figure}

Design~\#1, shown in the top panel of Figure~\ref{fig:designs}, has a
curved fibre slit with a fish-tail design that sends the light from half
the fibres to one side of the collimator and half to the other. The
fibre apertures are approximately 1~arcsec. The two parallel collimated
beams, each of diameter 150\,mm are sent to two volume phase holographic
(VPH) gratings. Two articulated cameras view the VPH gratings in the
Littrow transmission configuration. Each camera has an F/2.3 dioptic
design and feeds an 8k$\times$8k CCD detector mosaic. Each of the two
spectrograph arms takes the light from 300 single fibres or 40 d-IFUs.

Design~\#2, shown in the bottom panel of Figure~\ref{fig:designs}, is
generally similar, except that the two parallel collimated beams are
each of diameter 300\,mm and the camera has an F/1.4 catadioptic design.
This allows the use of a smaller 4k$\times$4k CCD detector. Each of the
two spectrograph arms takes the light from 300 single fibres or 30
d-IFUs.

\section{Comparison with other VLT facilities}

MAXIMUS extends the capabilities of GIRAFFE by offering a wider range of
spectral resolutions: $R\sim1800$--30000 compared to GIRAFFE's
$R\sim5000$--20000. The lower resolutions in particular will make
MAXIMUS much more useful for extragalactic observations. MAXIMUS has the
same field area (0.136\,deg$^2$), but a 4.5$\times$ higher multiplex for
single fibres (600 vs 132). MAXIMUS also offers 60--80 deployable IFUs
each with a field of view of 2~arcsec diameter at 0.3~arcsec resolution,
compared to GIRAFFE's 15 deployable IFUs each with a field of view of
3~arcsec\,$\times$\,2~arcsec at 0.6~arcsec resolution. By using volume
phase holographic gratings, MAXIMUS should achieve 30\% higher
throughput than GIRAFFE (excluding detector differences).

MAXIMUS complements the capabilities of VIMOS by offering 10$\times$ the
spectral resolution ($R\sim1800$--30000 vs $R\sim200$--2500) and
2.2$\times$ the field area (0.136\,deg$^2$ vs 0.062\,deg$^2$). MAXIMUS
gives a similar multiplex to VIMOS at low resolution (600 vs 800) at
lower surface density (1.0\,deg$^{-2}$ vs 3.6\,deg$^{-2}$); at high
resolution it provides 3$\times$ higher multiplex (600 vs 200) at
comparable surface density (1.0\,deg$^{-2}$ vs 0.9\,deg$^{-2}$). The IFU
facilities offered are also complementary: for MAXIMUS, 60--80
deployable IFUs with field of view 2~arcsec diameter at 0.3~arcsec
resolution vs the single large VIMOS IFU.

\section{A near-infrared version of MAXIMUS}

A near-infrared version of MAXIMUS (design~\#3) can also be envisaged.
This design is similar to design~\#1, using F/2.3 dioptic cameras with
4k$\times$4k HgCdTe detectors that would be interchangeable with the
optical cameras. The individual object apertures would again be about
1~arcsec in diameter, but in this design would consist of a 7-hex-format
fibre bundle, allowing about 250 single objects or 20 deployable IFUs to
be accommodated in each spectrograph.

This NIR version of MAXIMUS would significantly extend the capabilities
of NIRMOS. It would offer 3.2$\times$ the spectral resolution
($R\sim8000$ vs $R\sim2500$), guaranteeing effective avoidance of the OH
sky lines. It would also have a 2.2$\times$ larger field area
(0.136\,deg$^2$ vs 0.062\,deg$^2$) and 2.8$\times$ the multiplex at
moderate resolution (500 vs 180). It would in addition offer 40
deployable IFUs, each with a field of view of 2~arcsec diameter at
0.3~arcsec sampling. Both MAXIMUS and NIRMOS would have the same
long-wavelength limit of around 1.6$\umu$m due to their non-cryogenic
design.

\end{document}